%% file: XYZ.tex
\documentclass[12pt]{article}
\usepackage{graphicx}
\def\pbnr{}
\def\speaker{Chang-Zheng Yuan}
\def\onbehalfof{}
\def\title{New results on XYZ states from $e^+e^-$ experiments }
\def\affiliation{Institute of High Energy Physics\\
Chinese Academy of Sciences, Beijing 100049, China}
\def\support{This work was supported
in part by the Ministry of Science and Technology of China under
Contract No. 2009CB825203, and National Natural Science Foundation
of China (NSFC) under Contracts Nos. 10825524, 10935008, and
11235011.}

\input charmmacros.tex
\newcommand{\y}{Y(4260)}
\newcommand{\z}{Z_c(3900)}
\newcommand{\x}{X(3872)}
\newcommand{\pp}{\pi^+\pi^-}
\newcommand{\ks}{K_S^0}

\newcommand{\LL}{\ell^+\ell^-}
\newcommand{\EE}{e^+e^-}
\newcommand{\MM}{\mu^+\mu^-}

\newcommand{\pphc}{\pi^+\pi^- h_c}

\newcommand{\psip}{\psi(3686)}

\newcommand{\jpsi}{J/\psi}
\newcommand{\piz}{\pi^0}

\newcommand{\zc}{Z_c(3900)}
\newcommand{\zcp}{Z_c(4020)}

\newcommand{\ppjpsi}{\pi^+\pi^-J/\psi}
\def\Journal#1#2#3#4{{#1} {\bf #2}, #3 (#4)}

\def\PLB{Phys. Lett. B}
\def\PRL{Phys. Rev. Lett.}
\def\PRD{Phys. Rev. D}

\def\EPJC{Eur. Phys. J. C}

\begin{document}
\begin{titlepage}
\pubblock

\vfill
\Title{\title}
\vfill
\Author{\speaker\SupportedBy{\support}\OnBehalf{\onbehalfof}}
\Address{\affiliation}
\vfill
\begin{Abstract}

In this talk, we present the recent study on the charmoniumlike
states from the $e^+e^-$ colliders, including BESIII, Belle,
BaBar, and CLEOc. The talk covers the $X(3872)$ from the $Y(4260)$
radiative transition, the $Y$ states from the
initial-state-radiation processes and from $\EE\to \pphc$, and the
charged $Z_c$ states.

\end{Abstract}
\vfill
\begin{Presented}
\venue
\end{Presented}
\vfill
\end{titlepage}
\def\thefootnote{\fnsymbol{footnote}}
\setcounter{footnote}{0}
%

\section{Introduction}

In the quark model, mesons are composed from one quark and one
anti-quark, while baryons are composed from three quarks. Although
no solid calculation shows hadronic states with other
configurations must exist in QCD, people believe hadrons with no
quark (glueball), with excited gluon (hybrid), or with more than
three quarks (multi-quark state) should exist. Since a proton and
a neutron can be bounded to form a deuteron, it is also believed
that other mesons can also be bounded to form molecules.

It is a long history of searching for all these kinds of states,
however, no solid conclusion was reached until recently on the
existence of any one of them, except deuteron.

At the $B$-factories, BaBar and Belle, lots of new states (called
charmoniumlike states \index{charmoniumlike states} or XYZ
particles \index{XYZ particles}) have been observed in the final
states with a charmonium and some light hadrons. All these states
populate in the charmonium mass region. They could be candidates
for charmonium states, however, there are also strange properties
shown from these states, these make them more like exotic states
rather than conventional mesons~\cite{epjc-review}. The
BESIII~\cite{bes3} experiment at the BEPCII collider started data
taking since 2009, and lots of data were accumulated at the peak
of the narrow vector charmonium resonances as well as above 4~GeV,
these data make the study of the XYZ states possible.

In this talk, we present the most recent results on the study of
the $X(3872)$ from the $\y$ radiative transition, the $Y$ states
from initial-state-radiation (ISR) processes and from $\EE$
annihilation, and the charged $Z_c$ states. The results are from
the BESIII~\cite{bes3}, Belle, BaBar, and CLEOc experiments.

\section{\boldmath Observation of the $\y\to \gamma\x$~\cite{BES3x}}

The $\x$ was observed by Belle in $B^\pm\to K^\pm\ppjpsi$ decays
ten years ago~\cite{bellex}. It was confirmed subsequently by
several other experiments~\cite{CDFx,D0x,babarx}. Since its
discovery, the $\x$ state has stimulated special interest for its
nature. Both BaBar and Belle observed $\x\to \gamma\jpsi$ decay,
which supports $\x$ being a C-even
state~\cite{babar-jpc,belle-jpc}. The CDF and LHCb experiments
determined the spin-parity of the $\x$ being
$J^{P}=1^{+}$~\cite{CDF-jpc,LHCbx}, and CDF experiment also found
that the $\pp$ system was dominated by a $\rho^0(770)$
resonance~\cite{CDF-pp}.

Being near $D\bar{D^*}$ mass threshold, the $\x$ was interpreted
as a good candidate for a hadronic molecule or a tetraquark
state~\cite{epjc-review}. Currently, $\x$ was only observed in B
meson decays and hadron collisions. Since the quantum number of
$\x$ is $J^{PC}=1^{++}$, it could be produced through the
radiative transition of the excited vector charmonium or
charmoniumlike states such as the $\psi$s and the $Y$s.

BESIII reported the observation of $\EE\to \gamma\x\to \gamma
\ppjpsi$, with $\jpsi$ reconstructed through its decays into
lepton pairs ($\LL=\EE$ or $\MM$). The analysis is performed with
the data samples collected with the BESIII detector taken at $\EE$
central-of-mass (CM) energies from $\sqrt{s}=4.009$~GeV to
4.420~GeV~\cite{zc4020}.

\begin{figure}
\begin{center}
\includegraphics[height=7cm]{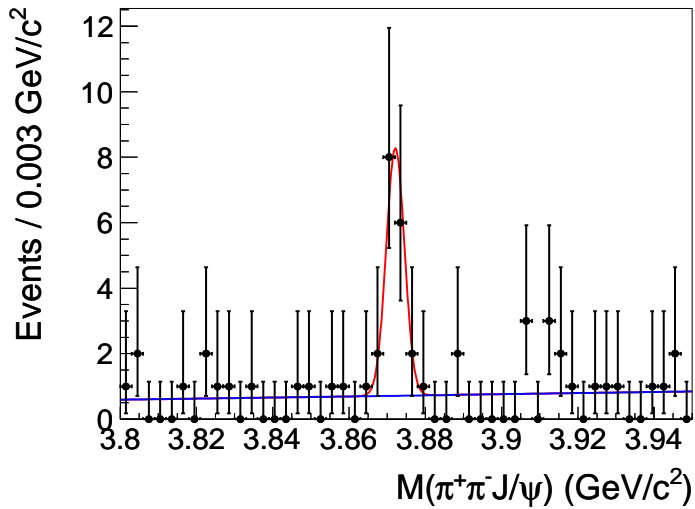}
\caption{Fit the $M(\ppjpsi)$ distribution with MC simulated
histogram convolving a Gaussian function for signal and a linear
background term. Dots with error bars are data, the curves are the
best fit.} \label{fit-mx}
\end{center}
\end{figure}

The $M(\ppjpsi)$ distribution (summed over all energy points) is
fitted to extract the mass and signal yield of the $\x$. An Monte
Carlo (MC) simulated signal histogram convolving a Gaussian
function which represents the difference between data and MC
simulation is taken as the signal shape, and a linear term is used
for the background. The ISR $\psip$ signal is used to calibrate
the absolute mass scale and to extract the resolution difference
between data and MC simulation. Figure~\ref{fit-mx} shows the fit
result, the measured mass of $\x$ is $M[\x] = (3872.1\pm 0.8\pm
0.3)$~MeV/$c^2$. The statistical significance of $\x$ is
$5.3\sigma$, estimated by comparing the difference of
log-likelihood value with and without $\x$ signal in the fit, and
taking the change of number-of-degree-of-freedom into
consideration.

The Born-order cross section is calculated using
 $
\sigma^{B}=\frac{N^{\rm obs}} {\mathcal{L}_{\rm int} (1+\delta)
\epsilon \mathcal{B}},
 $
where $N^{\rm obs}$ is the number of observed events obtained from
the fit to the $M(\ppjpsi)$ distribution, $\mathcal{L}_{\rm int}$
is integrate luminosity, $\epsilon$ is selection efficiency,
$\mathcal{B}$ is branching ratio of $\jpsi\to \LL$ and
($1+\delta$) is the radiative correction factor. The results are
listed in Table~\ref{sec}. For 4.009~GeV and 4.360~GeV data, since
the $\x$ signal is not significant, upper limits on the production
rates are given at 90\% C.L.

\begin{table*}
\begin{center}
\caption{The product of the Born cross section $\sigma^{B}(\EE\to
\gamma \x)$ and $\mathcal{B}(\x\to \ppjpsi)$ at different energy
points. The upper limits are given at 90\% C.L.} \label{sec}
\begin{tabular}{cc}
  \hline\hline
  $\sqrt{s}$~(GeV)  & $\sigma^B[\EE\to \gamma\x]\cdot\mathcal{B}(\x\to \ppjpsi)$~(pb) \\  \hline
  4.009 &  $<0.12$  \\
  4.229 &  $0.32\pm 0.15\pm 0.02$  \\
  4.260 &  $0.35\pm 0.12\pm 0.02$  \\
  4.360 &  $<0.39$ \\
  \hline\hline
\end{tabular}
\end{center}
\end{table*}

The observation suggests that the $\x$ might be from the radiative
transition of the $\y$. Combining with the $\EE\to \ppjpsi$ cross
section measurement at $\sqrt{s}=4.260$~GeV from
BESIII~\cite{zc3900}, one obtains $\sigma^B[\EE\to \gamma\x]\cdot
\mathcal{B}[\x\to \ppjpsi]/\sigma^B(\EE\to \ppjpsi) = (5.6\pm
2.0)\times 10^{-3}$, under the assumption that $\x$ produced only
from $\y$ radiative decays. If one takes $\mathcal{B}[\x\to
\ppjpsi]=5\%$~\cite{bnote}, then $\mathcal{R} =
\frac{\sigma^B[\EE\to \gamma\x]}{\sigma^B(\EE\to \ppjpsi)}\sim
11\%$. The measured relative large production rate near 4.26~GeV
shows a similar tendency to the model dependent
calculation~\cite{model}.

\section{More information on the $Y$ states}

The study of charmonium states via ISR at the $B$-factories has
proven to be very fruitful. In the process $e^+e^- \to \gamma_{\rm
ISR} \pi^+\pi^-J/\psi$, the BaBar experiment observed the
$Y(4260)$~\cite{babary}. This structure was also observed by the
CLEO~\cite{cleoy} and Belle experiments~\cite{belley} with the
same technique; moreover, there is a broad structure near
4.008~GeV in the Belle data. In an analysis of the $e^+e^- \to
\gamma_{\rm ISR} \pi^+\pi^-\psi(2S)$ process, BaBar found a
structure at around 4.32~GeV~\cite{babar_pppsp}, while the Belle
observed two resonant structures at 4.36~GeV and
4.66~GeV~\cite{belle_pppsp}. Recently, BaBar updated $e^+e^- \to
\gamma_{\rm ISR} \pi^+\pi^-\psi(2S)$ analysis with the full data
sample, and confirmed the $Y(4660)$ state~\cite{babar_pppsp_new};
while the update of the $e^+e^- \to \gamma_{\rm ISR}
\pi^+\pi^-J/\psi$ from both the BaBar and Belle experiments still
show differences at the $Y(4008)$ mass
region~\cite{babary_new,belley_new}.

BESIII experiment reported the cross section of $\EE\to \pp h_c$
final state with 13 energy points between 3.81 and
4.42~GeV~\cite{zc4020}, together with the CLEOc measurement at
4.17~GeV~\cite{cleoc_pipihc}, the data indicate the existence of a
narrow structure at around 4.22~GeV.

\subsection{\boldmath Confirmation of the $Y(4660)$~\cite{babar_pppsp_new}}

The BaBar experiment study the process $e^+e^-\to \pp\psi(2S)$
with ISR events. The data were recorded with the BaBar detector at
center-of-mass energies at and near the $\Upsilon(\mathrm{nS})$ (n
= 2, 3, 4) resonances and correspond to an integrated luminosity
of 520~fb$^{-1}$. They investigate the $\pp\psi(2S)$ mass
distribution from 3.95 to 5.95~GeV, and measure the center-of-mass
energy dependence of the associated $e^+e^-\to \pp\psi(2S)$ cross
section. The mass distribution exhibits evidence of two resonant
structures. A fit to the $\pp\psi(2S)$ mass distribution
corresponding to the decay mode $\psi(2S)\to \pp J/\psi$ yields a
mass value of $4340\pm 16\pm 9$~MeV/$c^2$ and a width of $94\pm
32\pm 13$~MeV for the $Y(4360)$, and for the $Y(4660)$ a mass
value of $4669\pm 21\pm 3$~MeV/$c^2$ and a width of $104\pm 48\pm
10$~MeV. The results are in good agreement with the Belle
measurement~\cite{belle_pppsp} and confirm the $Y(4660)$ observed
by the Belle experiment.

\subsection{\boldmath Measurement of $\EE\to \pp h_c$}

BESIII studied $\EE\to \pp h_c$ at 13 CM energies from 3.900 to
4.420~GeV~\cite{zc4020}. The data samples and the results are
listed in Table~\ref{scan-data}. In the studies, the $h_c$ is
reconstructed via its electric-dipole (E1) transition $h_c\to
\gamma\eta_c$ with $\eta_c\to X_i$, where $X_i$ signifies 16
exclusive hadronic final states: $p \bar{p}$, $2(\pi^+ \pi^-)$,
$2(K^+ K^-)$, $K^+ K^- \pi^+ \pi^-$, $p \bar{p} \pi^+ \pi^-$,
$3(\pi^+ \pi^-)$, $K^+ K^- 2(\pi^+ \pi^-)$, $\ks K^\pm \pi^\mp$,
$\ks K^\pm \pi^\mp \pi^\pm \pi^\mp$, $K^+ K^- \pi^0$, $p
\bar{p}\pi^0$, $\pi^+ \pi^- \eta$, $K^+ K^- \eta$, $2(\pi^+ \pi^-)
\eta$, $\pi^+ \pi^- \pi^0 \pi^0$, and $2(\pi^+ \pi^-) \pi^0
\pi^0$. Here $K_S^0$ is reconstructed from its $\pi^+\pi^-$
decays, and the $\piz$ and $\eta$ from their $\gamma \gamma$ final
states.

\begin{table}[htbp]
\caption{$\EE\to \pphc$ cross sections (or upper limits at the
90\% confidence level). The third errors are from the uncertainty
in ${\cal B}(h_c\to \gamma\eta_c)$~\cite{bes3-hc-inclusive}.}
\label{scan-data} \centering
\begin{tabular}{crcccc}
  \hline\hline
  $\sqrt{s}$~(GeV) & ${\cal L}$ (pb$^{-1}$)
  & $n^{\rm obs}_{h_c}$ & $\sigma(\EE\to \pphc)$~(pb) \\
  \hline
  3.900  &  52.8~~ & $<2.3$  & $<8.3 $ \\
  4.009  & 482.0~~ & $<13$  & $<5.0$ \\
  4.090  &  51.0~~ & $<6.0$  & $<13$ \\
  4.190  &  43.0~~ & $8.8\pm 4.9$  & $17.7\pm  9.8\pm  1.6\pm 2.8$ \\
  4.210  &  54.7~~ & $21.7\pm 5.9$ & $34.8\pm  9.5\pm  3.2\pm 5.5$ \\
  4.220  &  54.6~~ & $26.6\pm 6.8$ & $41.9\pm 10.7\pm  3.8\pm 6.6$ \\
  4.230  &1090.0~~ & $646\pm 33$   & $50.2\pm  2.7\pm  4.6\pm 7.9$ \\
  4.245  &  56.0~~ & $22.6\pm 7.1$ & $32.7\pm 10.3\pm  3.0\pm 5.1$ \\
  4.260  & 826.8~~ & $416\pm 28$   & $41.0\pm  2.8\pm  3.7\pm 6.4$ \\
  4.310  &  44.9~~ & $34.6\pm 7.2$ & $61.9\pm 12.9\pm  5.6\pm 9.7$ \\
  4.360  & 544.5~~ & $357\pm 25$   & $52.3\pm  3.7\pm  4.8\pm 8.2$ \\
  4.390  &  55.1~~ & $30.0\pm 7.8$ & $41.8\pm 10.8\pm  3.8\pm 6.6$ \\
  4.420  &  44.7~~ & $29.1\pm 7.3$ & $49.4\pm 12.4\pm  4.5\pm 7.6$ \\
  \hline\hline
\end{tabular}
\end{table}

\begin{figure}[htbp]
\centering
 \includegraphics[width=10cm]{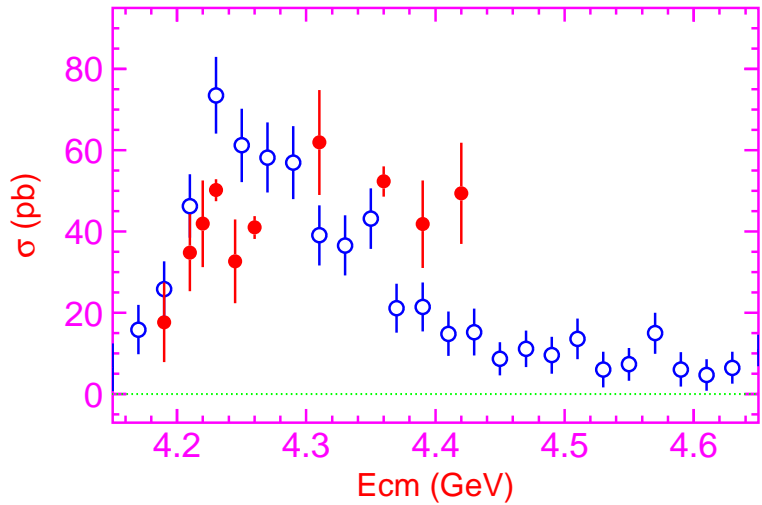}
\caption{The comparison between the cross sections of $\EE\to \pp
h_c$ from BESIII~(dots with error bars) and those of $\EE\to
\ppjpsi$ from Belle~(open circles with error bars). The errors are
statistical only.}
  \label{pphc_ppjpsi}
\end{figure}

The resulting cross sections are of the same order of magnitude as
those of the $\EE\to \ppjpsi$ measured by BESIII~\cite{zc3900} and
other experiments~\cite{babary_new,belley_new}, but with a
different line shape (see Fig.~\ref{pphc_ppjpsi}). There is a
broad structure at high energy with a possible local maximum at
around 4.23~GeV. Together with the measurement at 4.17~GeV by the
CLEOc experiment~\cite{cleoc_pipihc}, $\sigma=(15.6\pm 2.3\pm
1.9\pm 3.0)$~pb, the cross sections are fit to extract the
possible resonant structures.

\begin{figure}[htbp]
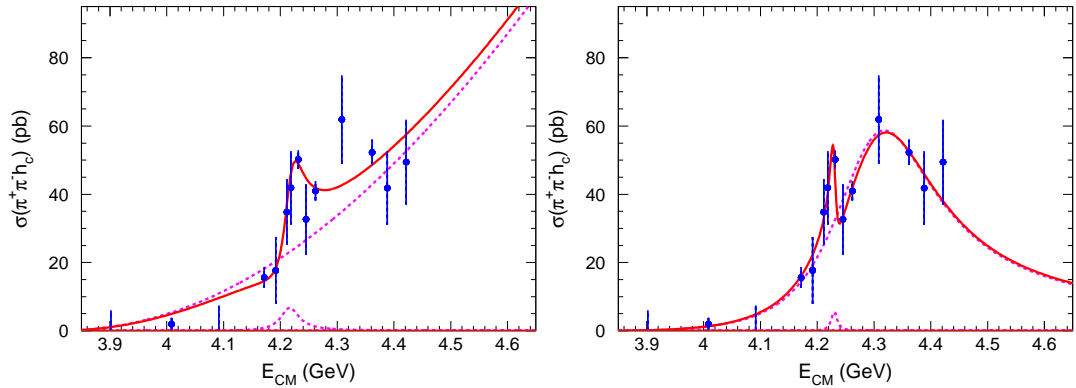

\centering
 \includegraphics[height=7cm,angle=-90]{BW+PS.epsi}
 \includegraphics[height=7cm,angle=-90]{2BW.epsi}
 \caption{The fit to the cross sections of $\EE\to \pp
h_c$ from BESIII and CLEOc~(dots with error bars). Solid curves
show the best fits, and the dashed ones are individual component.
Left panel is the fit with the coherent sum of a phase space
amplitude and a BW function, and the right panel is the coherent
sum of two BW functions.}
  \label{fit_pipihc}
\end{figure}

As the line shape at above 4.42~GeV is unknown, it is not clear
whether the large cross section at high energy will decrease or
not. Assuming the cross section follows the three-body phase space
and there is a narrow resonance at around 4.2~GeV, we fit the
cross sections with the coherent sum of two amplitudes, a constant
and a constant width relativistic Breit-Wigner (BW) function. The
fit indicates the existence of a resonance (called $Y(4220)$
hereafter) with a mass of $(4216\pm 7)$~MeV/$c^2$ and width of
$(39\pm 17)$~MeV, and the goodness-of-the-fit is $\chi^2/{\rm ndf}
= 11.04/9$, corresponding to a confidence level of 27\%. There are
two solutions for the $\Gamma_{\EE}\times {\cal B}(Y(4220)\to
\pphc)$ which are $(3.2\pm 1.5)$~eV and $(6.0\pm 2.4)$~eV. Fitting
the cross sections without the $Y(4220)$ results in a very bad
fit, $\chi^2/{\rm ndf} = 72.75/13$, corresponding to a confidence
level of $2.5\times 10^{-10}$. The statistical significance of the
$Y(4220)$ is calculated to be $7.1\sigma$ comparing the two
$\chi^2$s obtained above and taking into account the change of the
number-of-degree-of-freedom. Figure~\ref{fit_pipihc}~(left panel)
shows the final fit with the $Y(4220)$.

Assuming the cross section decreases at high energy, we fit the
cross sections with the coherent sum of two constant width
relativistic BW functions. The fit indicates the existence of the
$Y(4220)$ with a mass of $(4230\pm 10)$~MeV/$c^2$ and width of
$(12\pm 36)$~MeV, as well as a broad resonance, the $Y(4290)$,
with a mass of $(4293\pm 9)$~MeV/$c^2$ and width of $(222\pm
67)$~MeV. The goodness-of-the-fit is $\chi^2/{\rm ndf} = 1.81/7$,
corresponding to a confidence level of 97\%. There are two
solutions for the $\Gamma_{\EE}\times {\cal B}(Y(4220)/Y(4290)\to
\pphc)$ which are $(0.07\pm 0.07)$~eV/$(16.1\pm 2.2)$~eV and
$(2.7\pm 4.9)$~eV/$(19.0\pm 5.9)$~eV. Fitting the cross sections
without the $Y(4220)$ results in a much worse fit, $\chi^2/{\rm
ndf} = 30.65/11$, corresponding to a confidence level of
$1.3\times 10^{-3}$. The statistical significance of the $Y(4220)$
is calculated to be $4.5\sigma$ comparing the two $\chi^2$s
obtained above and taking into account the change of the
number-of-degree-of-freedom. Figure~\ref{fit_pipihc}~(right panel)
shows the final fit with the $Y(4220)$ and $Y(4290)$.

From the two fits showed above, we conclude that there must be a
narrow structure at around 4.22~GeV/$c^2$ with a width at
10$-$50~MeV level, although we are not sure if there is a broad
resonance at 4.29~GeV/$c^2$. More measurements from the BESIII
experiments at CM energies above 4.42~GeV will certainly tell
which of the above two fits is more meaningful, and more precise
data at around the $Y(4220)$ peak will also be crucial to extract
the resonant parameters of it more precisely.

There are thresholds of $\bar{D}D_1$~\cite{zhaoq},
$\omega\chi_{cJ}$~\cite{zhenghq,yuancz}, $D_s^{\ast+}D_s^{\ast-}$
at the $Y(4220)$ mass region, these make the identification of the
nature of this structure very complicated. It is worth to point
out that the lattice QCD calculations indicate that the
charmonium-hybrid lies in the mass region of these two $Y$
states~\cite{ccg_lqcd} and the $c\bar{c}$ tend to be in a
spin-singlet state. Such a state may couple to a spin-singlet
charmonium state such as $h_c$ strongly, this makes the $Y(4220)$
and/or $Y(4290)$ good candidates for the charmonium-hybrid states.

\section{Observation of charged charmoniumlike states}

In the study of $\EE\to \ppjpsi$ at CM energies around 4.26~GeV,
the BESIII~\cite{zc3900} and Belle~\cite{belley_new} experiments
observed a charged charmoniumlike state, the $\zc$, which was
confirmed shortly after with CLEO data at a CM energy of
4.17~GeV~\cite{seth_zc}. As there are at least four quarks within
the $\zc$, it is interpreted either as a tetraquark state,
$D\bar{D^*}$ molecule, hadro-quarkonium, or other configuration.
More recently, BESIII observed a charged $Z_{c}(4025)$ state in
$\EE\to \pi^\pm(D^*\bar{D}^*)^\mp$~\cite{zc4025} and a charged
$Z_{c}(4020)$ state in $\EE\to \pi^\pm(\pi^\mp
h_c)$~\cite{zc4020}. These states seem to indicate that a new
class of hadrons has been observed.

\subsection{\boldmath Observation of the $Z_c(3900)$~\cite{zc3900,belley_new}}

\begin{figure}[htbp]
 \includegraphics[height=5.3cm]{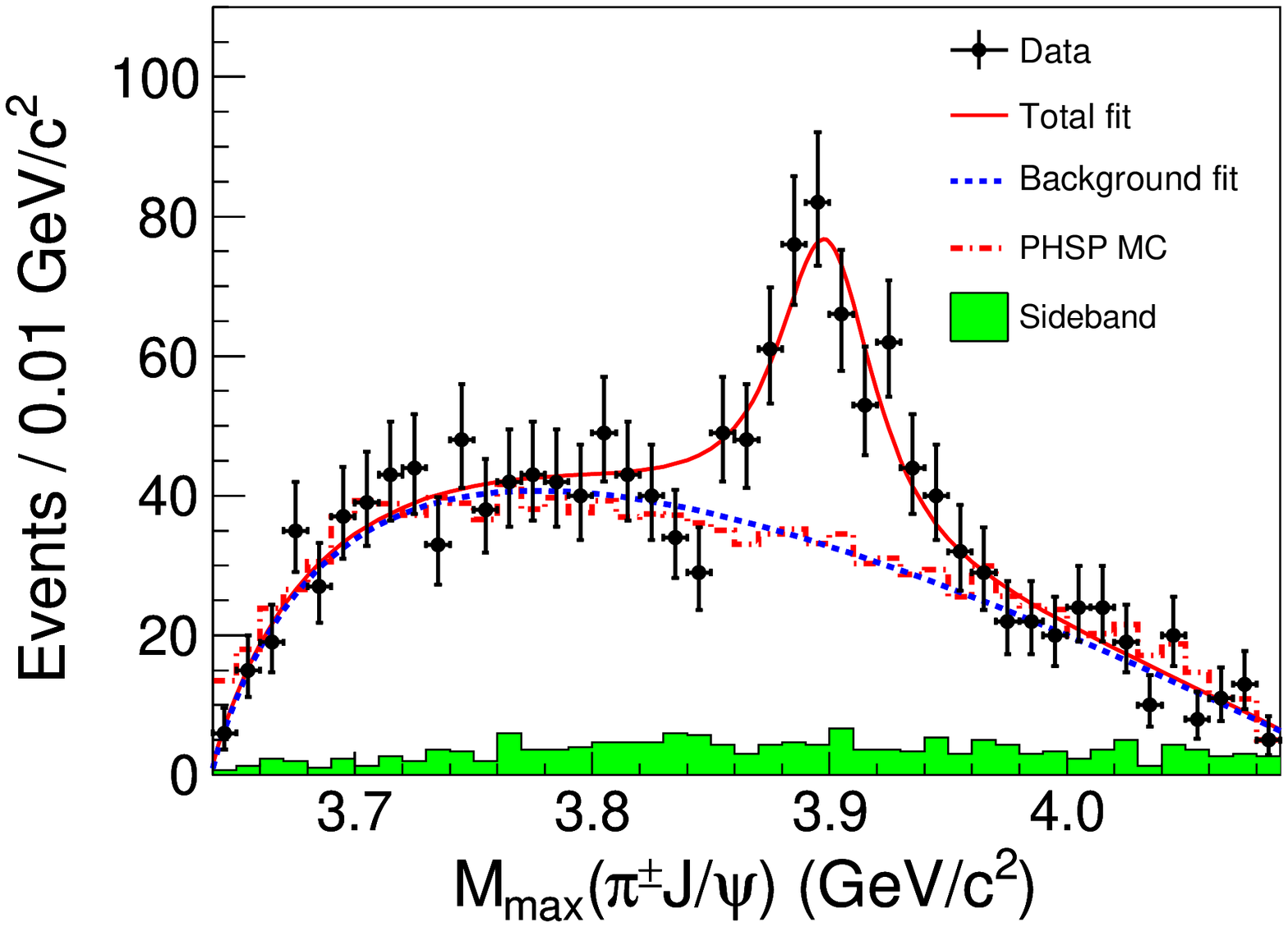}
 \includegraphics[height=5.3cm]{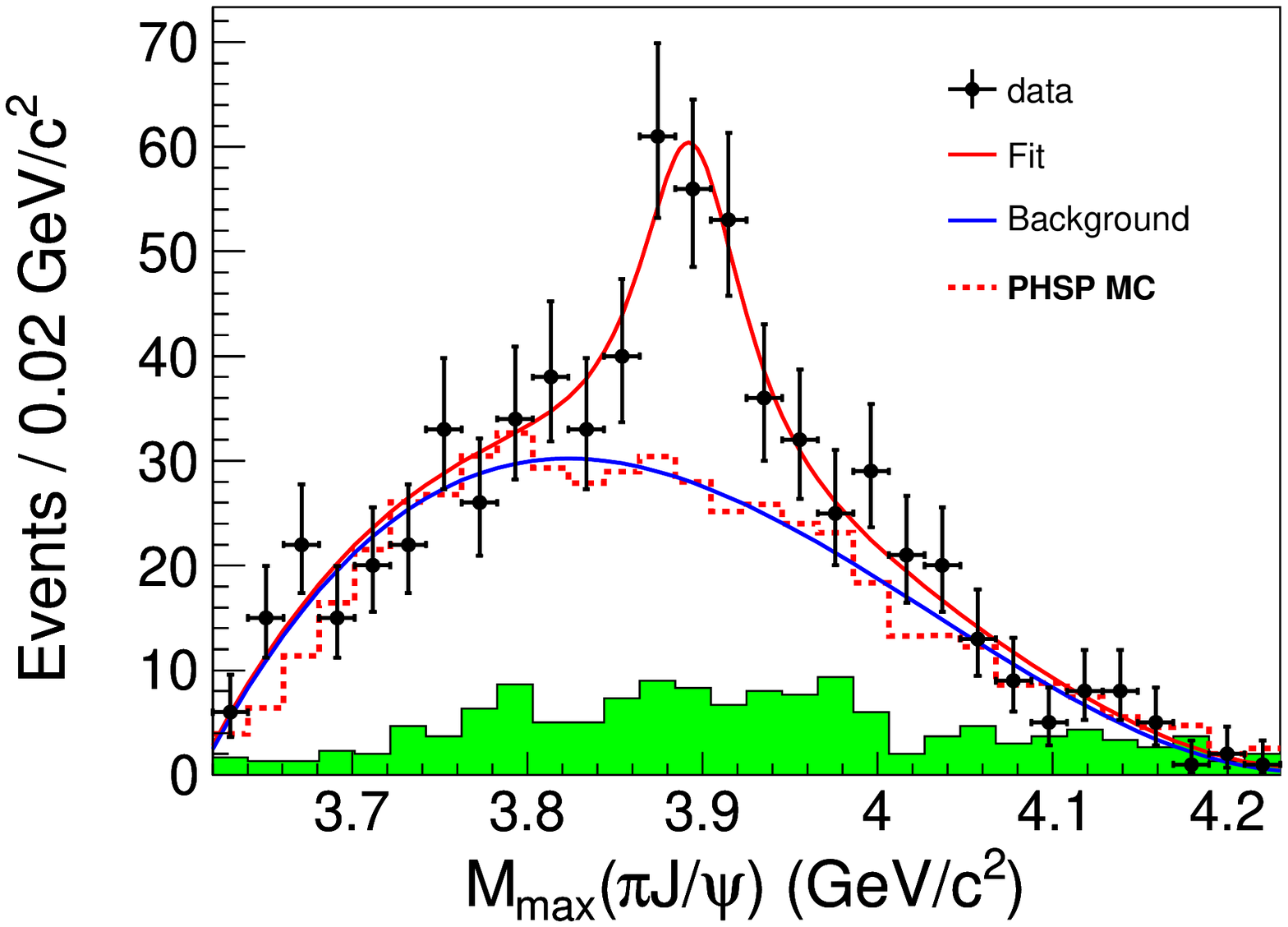}
 \caption{Unbinned maximum likelihood fit to the distribution of
the $M_{\mathrm{max}}(\pi J/\psi)$ (left panel from BESIII and
right panel from Belle). Points with error bars are data, the
curves are the best fit, the dashed histograms are the phase space
distributions and the shaded histograms are the non-$\ppjpsi$
background estimated from the normalized $\jpsi$ sidebands.}
\label{projfit}
\end{figure}

BESIII experiment studied the process $\EE\to \ppjpsi$ at a CM
energy of $4.260$~GeV using a 525~pb$^{-1}$ data
sample~\cite{zc3900}. A structure at around 3.9~GeV/$c^2$ is
observed in the $\pi^\pm \jpsi$ mass spectrum with a statistical
significance larger than $8\sigma$, which is referred to as the
$\zc$. If interpreted as a new particle, it is unusual in that it
carries an electric charge and couples to charmonium. A fit to the
$\pi^\pm\jpsi$ invariant mass spectrum (see Fig.~\ref{projfit}),
neglecting interference, results in a mass of $(3899.0\pm 3.6\pm
4.9)~{\rm MeV}/c^2$ and a width of $(46\pm 10\pm 20)$~MeV. Its
production ratio is measured to be $R=\frac{\sigma(\EE\to \pi^\pm
\zc^\mp\to \ppjpsi))} {\sigma(\EE\to \ppjpsi)}=(21.5\pm 3.3\pm
7.5)\%$.

At Belle experiment, the cross section of $\EE\to \ppjpsi$ is
measured from 3.8~GeV to 5.5~GeV using ISR method. The $\y$
resonance is observed and its resonant parameters are determined.
The intermediate states in $\y\to \ppjpsi$ decays are also
investigated~\cite{belley_new}. A $\zc$ state with a mass of
$(3894.5\pm 6.6\pm 4.5)~{\rm MeV}/c^2$ and a width of $(63\pm
24\pm 26)$~MeV/$c^2$ is observed in the $\pi^\pm\jpsi$ mass
spectrum (see Fig.~\ref{projfit}) with a statistical significance
larger than $5.2\sigma$.

The $\zc$ was confirmed shortly after with CLEOc data at a CM
energy of 4.17~GeV~\cite{seth_zc}, the mass and width agree with
the BESIII and Belle measurements very well.

This state is close to the $D\bar{D}^*$ mass threshold. As the
$\z$ state has a strong coupling to charmonium and is charged, it
cannot be a conventional $c\bar{c}$ state. However, its nature is
unknown.

\subsection{\boldmath Observation of the $Z_c(4020)$ and $Z_c(4025)$}

BESIII experiment measured $\EE\to \pphc$ cross sections at CM
energies between 3.90 and 4.42~GeV and analyzed the Dalitz plot of
$\pphc$ system. A narrow structure very close to the
$(D^\ast\bar{D}^\ast)^\pm$ threshold with a mass of $(4022.9\pm
0.8\pm 2.7)~{\rm MeV}/c^2$ and a width of $(7.9\pm 2.7\pm
2.6)$~MeV is observed in the $\pi^\pm h_c$ mass
spectrum~\cite{zc4020}. This structure couples to charmonium and
has an electric charge, which is suggestive of a state containing
more quarks than just a charm and an anti-charm quark, as the
$\zc$ observed in the $\pi^\pm\jpsi$
system~\cite{zc3900,belley_new,seth_zc}. BESIII experiment does
not not find a significant signal for $\zc\to\pi^\pm h_c$ and the
production cross section is found to be smaller than 11~pb at the
90\% C.L. at 4.26~GeV, which is lower than that of $\zc\to
\pi^\pm\jpsi$~\cite{zc3900}.

\begin{figure}[htbp]
\begin{center}
\includegraphics[width=10cm]{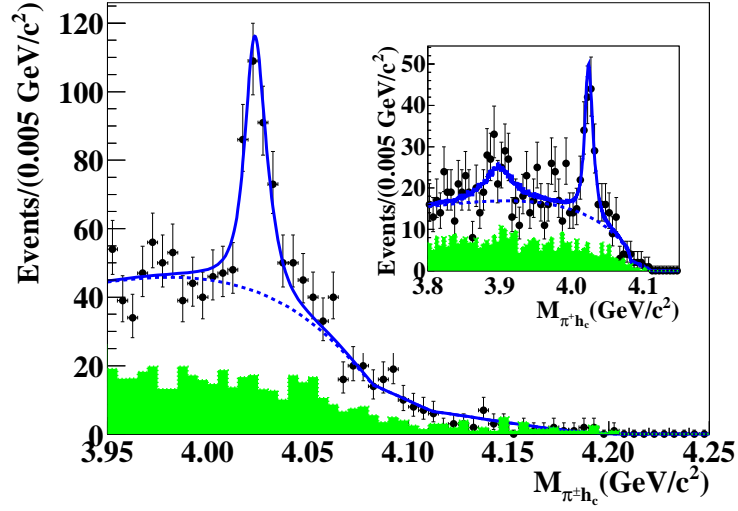}
\caption{Sum of the simultaneous fits to the $M_{\pi^\pm h_c}$
distributions at 4.23~GeV, 4.26~GeV, and 4.36~GeV; the inset shows
the sum of the simultaneous fit to the $M_{\pi^+ h_c}$
distributions at 4.23~GeV and 4.26~GeV with $\zc$ and $\zcp$. Dots
with error bars are data; shaded histograms are normalized
sideband background; the solid curves show the total fit, and the
dotted curves the backgrounds from the fit.} \label{1Dfit}
\end{center}
\end{figure}

BESIII experiment also studied the process $e^+e^- \to (D^{*}
\bar{D}^{*})^{\pm} \pi^\mp$ at a center-of-mass energy of 4.26~GeV
using a 827~pb$^{-1}$ data sample. Based on a partial
reconstruction technique, the Born cross section is measured to be
$(137\pm 9\pm 15)$~pb. A structure near the $(D^{*}
\bar{D}^{*})^{\pm}$ threshold in the $\pi^\mp$ recoil mass
spectrum is observed, which is denoted as the $Z_c(4025)$. The
measured mass and width of the structure are $(4026.3\pm 2.6\pm
3.7)$~MeV/$c^{2}$ and $(24.8\pm 5.6\pm 7.7)$~MeV, respectively,
from a fit with a constant width BW function for the signal. Its
production ratio $\frac{\sigma(e^+e^-\to Z^{\pm}_c(4025)\pi^\mp
\to (D^{*} \bar{D}^{*})^{\pm} \pi^\mp)}{\sigma(e^+e^-\to (D^{*}
\bar{D}^{*})^{\pm} \pi^\mp)}$ is determined to be $0.65\pm 0.09\pm
0.06$.

\begin{figure}[htbp]
\centering
\includegraphics[width=8cm]{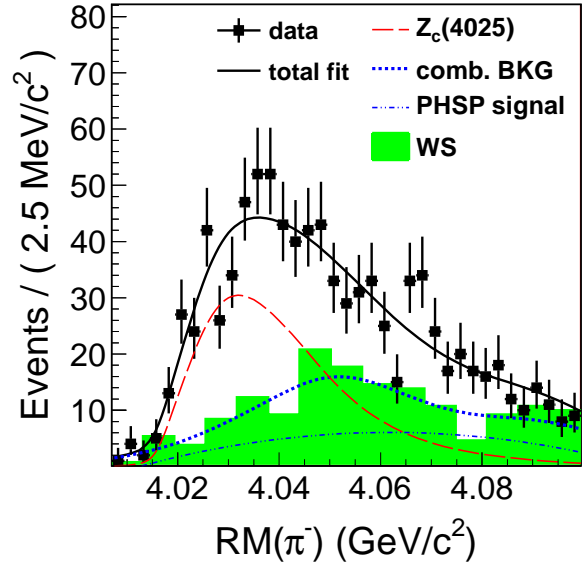}
\caption{Unbinned maximum likelihood fit to the $\pi^\mp$ recoil
mass spectrum in $e^+e^- \to (D^{*} \bar{D}^{*})^{\pm} \pi^\mp$. }
\label{fig:fit}
\end{figure}

The $\zcp$ parameters agree within 1.5$\sigma$ of those of the
$Z_c(4025)$. Currently one cannot tell whether they are the same
state. Further study is needed.

\section{Summary}

In summary, there are lots of charmoniumlike states observed in
charmonium mass region but many of them show properties different
from the naive expectation of conventional charmonium states. The
BESIII experiment is now producing results on these XYZ states.
The observation of the charged charmonium states, $\zc$, $\zcp$,
and $Z_c(4025)$, may indicate one kind of the exotic states has
been observed.

In the near future, BESIII experiment~\cite{bes3} will accumulate
more data for further study; the Belle II experiment~\cite{belle2}
under construction, with about 50~ab$^{-1}$ data accumulated, will
surely improve our understanding of all these states.

\Acknowledgements

We thank the organizers for their invitation to give the review
talk and congratulate the organizers for a very successful
workshop.

\end{document}

%% file: charmmacros.tex
\newcommand\pubnumber{\pbnr}
\newcommand\pubdate{\today}
\def\Title#1{\begin{center} {\Large #1 } \end{center}}
\def\Author#1{\begin{center}{ \sc #1} \end{center}}

\newcommand{\OnBehalf}[1]{\sbox0{#1}\ifdim\wd0=0pt
        {}
	\else
	{\\on behalf of #1}
	\fi}
\newcommand{\SupportedBy}[1]{\sbox0{#1}\ifdim\wd0=0pt
        {}
	\else
	{\footnote{#1}}
	\fi}
\def\Address#1{\begin{center}{ \it #1} \end{center}}

\newcommand\pubblock{\includegraphics[width=5cm]{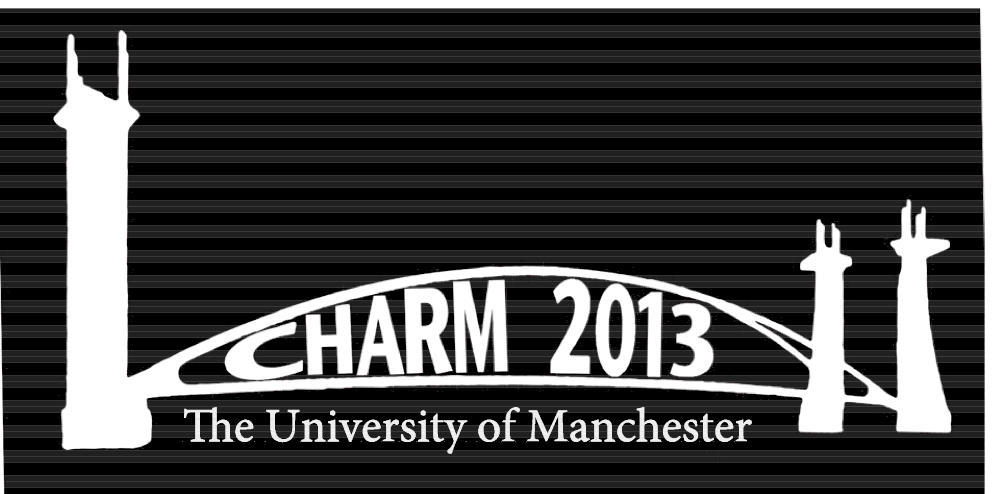}\hfill{\begin{tabular}{l} \pubnumber\\
         \pubdate  \end{tabular}}}
\newenvironment{Abstract}{\begin{quotation}  }{\end{quotation}}
\newenvironment{Presented}{\begin{quotation} \begin{center} 
             PRESENTED AT\end{center}\bigskip 
      \begin{center}\begin{large}}{\end{large}\end{center} \end{quotation}}
\def\Acknowledgements{\bigskip  \bigskip \begin{center} \begin{large}
             \bf ACKNOWLEDGEMENTS \end{large}\end{center}}
\def\venue{The 6$^{th}$ International Workshop on Charm Physics\\
(CHARM 2013)\\
Manchester, UK,  31 August -- 4 September, 2013}

\def\beq{\begin{equation}}
\def\eeq#1{\label{#1}\end{equation}}
\def\eeqn{\end{equation}}

\def\beqa{\begin{eqnarray}}
\def\eeqa#1{\label{#1}\end{eqnarray}}
\def\eeqan{\end{eqnarray}}

\let\bar=\overbar

\def\Dslash{\not{\hbox{\kern-4pt $D$}}}
\def\dslash{\not{\hbox{\kern-2pt $\del$}}}

\def\msb{{\bar{\ssstyle M \kern -1pt S}}}

